# A Framework for Undergraduate Data Collection Strategies for Student Support Recommendation Systems in Higher Education

Herkulaas MvE Combrink[1][0000-0001-7741-3418], Vukosi Marivate[1][0000-0002-6731-6267], and Benjamin Rosman[2][0000-0002-0284-4114]

1 University of Pretoria, Pretoria, RSA CombrinkHM@ufs.ac.za
  vukosi.marivate@cs.up.ac.za
2 University of the Witwatersrand, Johannesburg, RSA benjros@gmail.com

**Abstract.** Understanding which student support strategies mitigate dropout and improve student retention is an important part of modern higher educational research. One of the largest challenges institutions of higher learning currently face is the scalability of student support. Part of this is due to the shortage of staff addressing the needs of students, and the subsequent referral pathways associated to provide timeous student support strategies. This is further complicated by the difficulty of these referrals, especially as students are often faced with a combination of administrative, academic, social, and socio-economic challenges. A possible solution to this problem can be a combination of student outcome predictions and applying algorithmic recommender systems within the context of higher education. While much effort and detail has gone into the expansion of explaining algorithmic decision making in this context, there is still a need to develop data collection strategies Therefore, the purpose of this paper is to outline a data collection framework specific to recommender systems within this context in order to reduce collection biases, understand student characteristics, and find an ideal way to infer optimal influences on the student journey. If confirmation biases, challenges in data sparsity and the type of information to collect from students are not addressed, it will have detrimental effects on attempts to assess and evaluate the effects of these systems within higher education.

## 1   Introduction

Improving student retention and reducing student dropout is a major part of 21st century higher education research [8,38]. In addition to this, understanding minority groups, creating equitable education strategies and reforming curricula to be inclusive in nature form part of the basis of decoloniality seen within developing countries [33,37,41]. One of the primary challenges institutions of higher learning are currently facing is the scalability of the staff complement to address the needs of the student cohort [10,18,42]. This is further complicated by the complexity of referrals especially from an academic advising perspective as students are often faced with a combination of challenges that influence their academic journey [19,20]. Furthermore, the challenges students face are larger than previously thought because it includes psychosocial and socio-economic support, in addition to the academic support required [12,40]. To attempt a potential scalable strategy to address this, the use of digital technologies may be implemented to reach the student cohort. A possible solution to this problem can be a combination of student outcome predictions and applying recommender systems within the context of higher education because the problem inadvertently presents itself to the aforementioned computational tools [23]. Student outcome predictions refer to the academic outcome of students within this context, and the accompanying referral, which if implemented early enough can better support the student. Within the context of the possible solutions to explore in an attempt to address a computational tool or set of tools within higher education, collaborative filtering, demographic based filtering or utility based recommender systems may be applied to this context [34].

Therefore, the purpose of this paper is to outline a data collection framework specific to recommender systems within this context. In order to reduce collection biases, understand student characteristics, and find an ideal way to collect information that can infer optimal influences on the student journey, critical consideration should be given to the challenges and references from previous work in this field.

## 2   Background

According to Gadinger (2014), immense pressure is placed upon institutions to produce more graduates, and studies in higher education critically engaging with these issues may contribute toward student throughput rates. In order to understand the needs of students, the types of interventions to implement and which student support strategies have been the most successful, student success, student engagement, academic advising, student transition and capabilities of students have to be taken into consideration [38].

One potential strategy that may be incorporated within learning analytics in higher education to computationally assist with the academic advising space, relates to solutions within data science, specifically recommender systems. Recommender systems are platforms that provide a proposition or commendation to users around a set of items. These systems make use of information filtering as the primary methodology to exclude redundant criteria around an item, and include items that are similar to either the user's likes, dislikes, or interests. The application of these systems are widely used in marketing, online shopping and entertainment. Within the education domain, recommender systems provide a feasible solution to the problem of streamlining a few recommendations from multiple referral pathways for a single user [43]. This is especially significant within a context of universities where a multitude of data are generated and stored about the student, in a domain where the recommendation that is made needs to be personalized to the user. Recommender systems function with the premise that data and context are given within a system around two entities, the user (which in this context is the student) and the item (which refers to the intervention presented to the user) [31].

How the information is processed within the algorithm depends on how the input data are filtered. Several different kinds of algorithms exist within this context, including filtering on the basis of: ratings (a user rating about a specific item), demography (race, gender, age, etc.), content data (textual analysis of items rated by the user or multiple users), or item-based collaborative filtering [23].

In item-based collaborative filtering, the objective is to observe a collection of items, denoted by , that the active user, $s(u,a)$ has rated. The items and the ratings are then computed to how similar they are to the target item $i_j$ which is then selected from the k most similar items $i_1, i_2,...,i_k$, based on their corresponding similarities $s_{i1}, s_{i2},...,s_{ik}$. Collaborative filtering differs from prediction functions in that prediction functions are expressed numerically ($r(a,j)$), and are concerned with finding the anticipated opinion of a user ($u_a$) for a specific item ($i_j$) in a process referred to as individual scoring. The output of a successful recommender system is either in the form of a prediction or recommendation. This approach (prediction) in the context of users who are students, however, is counter intuitive within the context of this problem because the scoring is dependent on the user in an environment where the recommendations could be social, socio-economic or psychosocial in nature. Another reason why the prediction approach is counter intuitive is because it takes the mean value between users into consideration without including all of the factors that affect student success, which in turn might not calculate the actual differences between students only similarities based on a few factors. An example of this similarity measure is Pearson correlation-based [28]. This infers that there are several different types of approaches to take when both designing and recommender systems, with ontology based systems as a reference point to include multiple viewpoints within the algorithm design process. Ontology in this context can be defined as a formal, explicit specification of a shared conceptualization [13].

By this definition, an ontology approach by virtue is more inclusive of different ideas because it is fundamentally integrated in the algorithm design process. As promising as ontology-based recommender systems may be, challenges related to biases within their approach have proactively been discussed within the context of human centered approaches [13]. Recommender systems in education have been extensively studied, especially from the application of learning styles as a user rating to an educational intervention [14,22,27,36]. Several questionnaires have been developed for educational recommender systems using learning styles as item based ratings and user rating identifiers, with varying results [9]. The varying results experienced by multiple studies are related to the concept that learning styles in the education do-main have been debunked [2,21,26,29, 30,39]. However, despite the debunking of learning styles as a school of thought, there is still implementation of the use of learning styles in the field of education based recommender systems[11,15].

This raises a concern in this field whereby the dispute increases questionable concerns if the underlying data collection instruments are potentially fundamentally flawed in their school of thought. This further elevates disquiets in terms of implementing systems intended to assist the student population, if the required data collection instruments for these systems have not been adapted for this context. Even more so, this is particularly important when considering that learning path generation and subsequent evaluation strategies become contextually difficult if the required instruments do not meet the appropriate metric to be applied for educational recommender systems. Data collection biases are prominent features if the instruments themselves are not applicable for this context. According to Pohl (2004), several biases exist within the context of designing a human centered approach, including confirmation biases, selection biases, implicit biases and reporting biases, just on the virtue of human involvement. In addition to addressing the challenges of bias, fairness and equitability within the context of these technologies need to be considered [5].

The "fairness" component refers to inclusive educational practices that are driven by new fundamental systems in the education domain that focus on including the entire cohort, instead of a specific group within its practices across all pedagogical practices. The "equity" element refers to personalizing elements of the education system to each individual student so that the outcome, not the journey, may be standardized across different students. Therefore, this study focuses on creating a framework for the education domain specifically related to data collection strategies in reducing these challenges.

## 3    Methodology

In order to understand the different data collection stratagems that are applied to recommender systems in the context of higher education, a desk research methodology was employed. The methodology was chosen to draw strengths from previous works in education, eLearning and recommender systems within this domain. Desk research refers to secondary research conducted on the findings of prior research. The areas of interest in this study are specific to recommender systems and include the data collection strategies employed by previous studies and industry practices. As far possible, data collection strategies specific for education were used, but there were studies that were included in the desk research to strengthen the argument. Furthermore, different recommender data collection strategies are included so as to create a consolidated data collection framework. A theoretical underpinning is required to frame data collection plans to mitigate bias and introduce data collection approaches within this context. In total, three distinct criteria of user item scenarios were explored which relate to data collection considerations, data collection features types and domain specific context Table 1

**Table 1.** Classification and description for desktop investigation.

| Classification | Description |
|---|---|
| Data collection considerations | Outlining data collection strategies, user specific data collection contexts |
| Data collection features | Outlining considerations to take into account related to features within the data types collected |
| Data collection domain specific contexts | Outlining information related to the specific contexts prevalent within higher education |

For simplification of the information, the user remains the student in all instances whereas the item differs in terms of the output, information about the user or information related to the specific task performed. Therefore, the recommender should factor in that user preferences might be the result of retrospective evidence about the user as an entity, rather than what the user (in this case student) thinks they require to be successful within university at that specific moment. This methodological concept is supported within higher education, and is even more pronounced when factoring in student transition and the challenges that may arise within society [16].

## 4   Results and Discussion

Most traditional recommender systems use only one type of recommendation, such as a specific item rating from a set of users. However, as pertained from previous studies, for a recommendation in the learning and education space to be effective, the data collection strategies should be dynamic and able to be evaluated using different types of systems [3,4,35,44]. The dynamic nature of the data collection strategies refers to being able to collect information across various areas of the user experience. Additionally, different types of systems include collecting information from students' pre, during, and post an education activity. Dynamic data collection strategies for educational purposes should include a variety of simple and complex data, allow for dynamic data changes and add not only to the user and item data, but also the domain knowledge (Table 2). The addition of domain knowledge adds to the concept of fairness and equity in the education space [5]. In addition to this, data collection should include the ability to retrospectively study the information that was gathered.

Information collected in the education space related to the specific education recommender problem should be a combination of static and dynamic data that can be used on a variety of scenarios (Table 2). This implies that any recommender data collection strategy should include a variety of dynamic approaches that are inclusive of understanding the education as well as the recommender problem. This means that university leaving students should complete some type of retrospective evaluation survey on the basis of their student journey. This also applies to students reflecting at the end of each semester or module (Fig. 1).

**Table 2.** Classification and description for desktop investigation.

| Data collection consideration | Recommendations/Areas for future work | Sources |
|---|---|---|
| Collect simple and complex data and profile capabilities | Develop comprehensive data structures, include experimental validation processes, select most suitable rating estimation function | [3] |
| Linguistic pre-processing | Retrospective validation of first study (in their context) needs to be conducted | [44] |
| Allow users to opt in or out of data collection strategies | Current applications should include multidimensional criteria and more than just the user and item | [35] |
| Extend user profile, extend item profile, add context and domain knowledge | Intelligent recommender systems using multi context models must be tested on several different types of problems and incorporate different evaluation techniques | [4] |

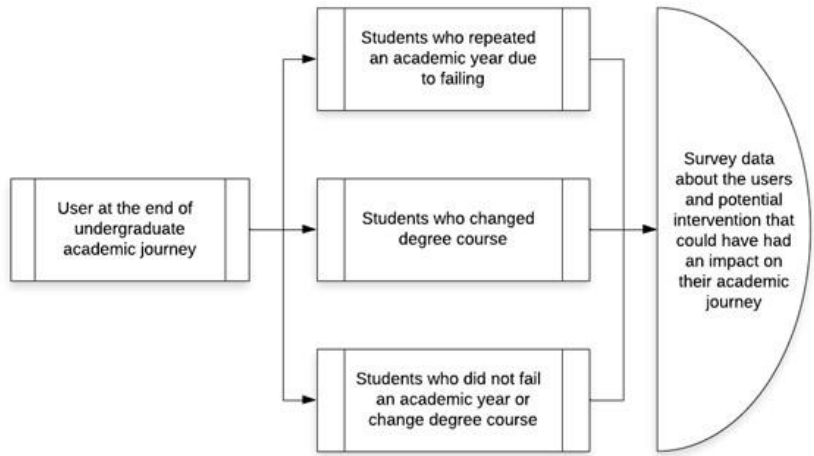

**Fig.1.** Data academic outcome metric.

However, it must be noted that the inclusion of retrospective evaluation is currently best positioned at the end of the academic journey, rather than at the end of each semester. Furthermore, users should be grouped into three categories, related to major changes in their student journey. These categories relate to students who repeated an academic year due to failing, students who changed degree course, and students who started and finished their initial degree without failing an academic year.

Research into the area of establishing fair algorithmic use so that users can under-stand the biases within recommendations in education has been explored (Table 3). According to Abdollahi and Nasraoui, (2018), "fair" machine learning models are inherently biased on the premise of the algorithm design, depending on the task. Moreover, the data and information that the algorithms are primed from in the con-text of the recommender problem are just as important as addressing bias filtering methods within the algorithm itself [1]. Ontology approaches have been explored in the education, eLearning, and recommender space, with an emphasis on various parts of the systems design, including an importance of data collection [31]. In 2014, the Open University published a "Policy on ethical use of student data for learning analytics" delimiting the nature and scope of data collected, emphasizing an explicit specification on the data that will not be collected and used for learning analytics [37]. This means that the ethics surrounding static data about the user (e.g. demography), dynamic data about the user (e.g. a change in interests), static and dynamic data about the domain (e.g. changes in institutional policy, faculty structures, etc.), and user rating and retrospective evaluation strategies about their own student path-way needs to be established. This is important because in general, evaluations within this specific domain are challenging unless a combination of user-item-domain data that is both static and dynamic is collected (Fig. 2). This further implies that institutional context plays a vital part in this journey.

**Table 3.** Data collection features, recommendations and sources.

| Data collection feature | Recommendations/Areas for future work | Sources |
|---|---|---|
| Data collection adds to the output of recommender systems in education | Ontology-based recommender systems require data collection strategies | [31] |
| Data collection frameworks are required in education | Policies should be implemented so that data collection strategies at a user level may be used in an ethical manner | [37] |
| The collection and dissemination strategies were interrogated | Include a user specific explanation to the users, incorporate multidimensional data collection strategies | [1] |

| A combination of static and dynamic data | Predict user state based on smartphone data and how to convey privacy measures implement-ed to user | [6] |
| --- | --- | --- |
| Applied to a static dataset | Design new metrics incorporating additional information related to user scores | [7] |
| Additional context around the data is required | Where do explanations about variable contribute toward the recommendation | [24] |
| Linguistic pre-processing | Retrospective validation of first study (in their context) needs to be conducted | [44] |

Scholarly engagement within the space of learning analytics may be limited, highlighting unique opportunities to engage in and understand students and the high-er education space better (Table 4). An attempt to understand the challenges students face, identifying contextual differences between students and institutions is on the Higher Education agenda, and grounded in erudite work within these fields. According to Kuh (2008), certain practices promote student success at pivotal moments within the curriculum and undergraduate experience. These interventions are related to high impact practices, but do not extend to social, psycho-social and socioeconomic support. With modernization and an increase in access to technologies popular within ma-chine learning applications, a possible solution to this problem may be rooted in data science. However, within this context several fallacies have been pointed out related to learning analytics within the student context [32].

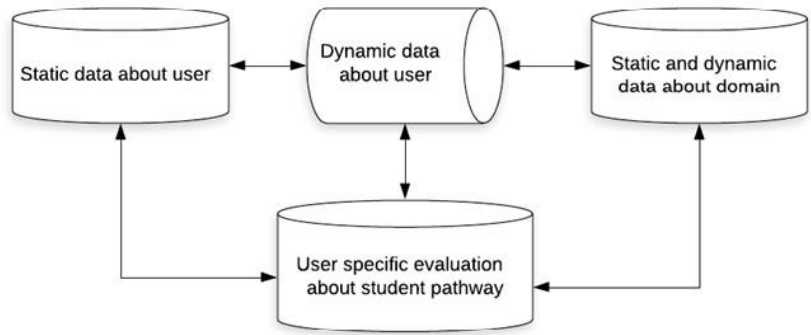

**Fig.2.** Data type differentiation.

These fallacies relate to isolated study design, circular reasoning with no real progress in addressing the problem and a lack of objectivity about the subject matter prior to implementation [32]. Based on their findings, and those before them in this domain, groundwork has been laid to address some of the fundamental programmatic biases that may arise within the understanding of the data that are collected in this setting [17]. To create an ethical and sound basis from which research may be conducted within the education space, data collection strategies about the domain contextualization is required (Table 4).

Within the education domain, student data collection can be grouped into first-year, intermediate, and final-year students. In order for a user to provide an accurate account of their journey, they need to be able to provide contextual information related to their perspective on academe, their field of study, and their university journey. Moreover, there are two streams of information required for each of the three areas of data collection. This relates to the hypothetical and actual account of the user's academic journey. The hypothetical account of the user journey is a retrospective reflective exercise in which the user rates a potential set of interventions that, in their opinion, could have assisted their academic journey. This refers to the actual account of their academic journey, including any academic interventions that occurred (Fig. 3).

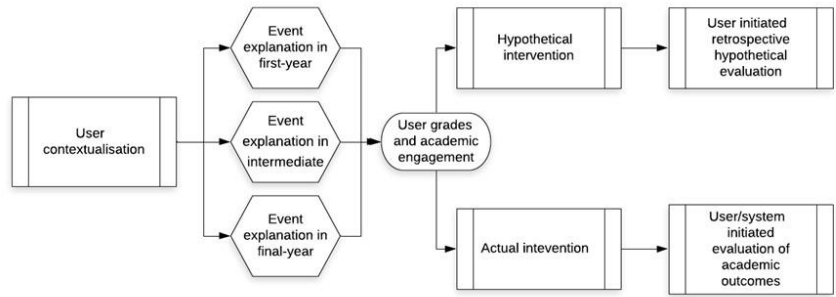

**Fig.3.** User academic journey contextualization.

In order to incorporate the recommendations proposed by the various authors that were identified in this study, we propose a detailed data collection framework (Fig. 4). This framework represents a consolidation of the groundwork set by previous studies, and consists of three sections related to data collection strategies, data collection variable types and domain specific context. The first succession and strategy within the framework is that data collection strategies must occur retrospectively about the student journey, as well as while the student journey is taking place. With the intention of creating a data collection strategy related to recommender systems in this domain, both the user pathway and the subsequent outcome is required. This is important for testing and evaluation strategies.

It must be emphasized that although data collection from the user is important, institutional data is also required in order to add to this framework to create a comprehensive dataset for the use of studying recommender systems in the higher education domain. This means that the data collection strategy should include a multidimensional data collection strategy from various areas of the higher education system from and by the students. A few potential instruments have been designed that are currently used in the learning analytics space that may be applied to this framework. This includes studies in student engagement, high impact practices and capabilities [25,38].

Furthermore, routinely collected institutional data should incorporated within these datasets in order to test various recommender system techniques within this domain. Lastly, all of the aforementioned conceptualizations need to be incorporated under the veil of ethics and in the context of studying human subjects [32].

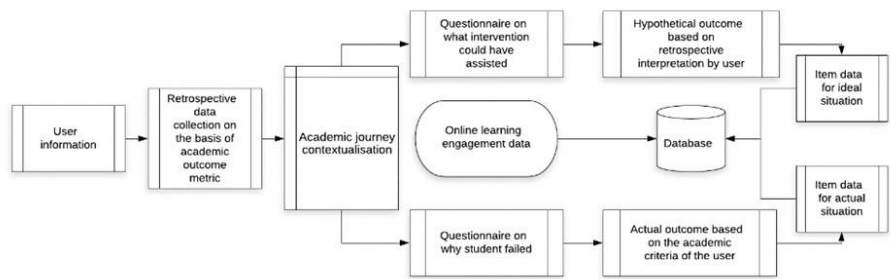

**Fig.4.** A framework for undergraduate data collection strategies for recommendation systems.

In addition to this, the proposed framework requires a representation of the student population and student demography profile, implying that a further investigation is required in order to incorporate fairness in sampling [1].

## 5   Conclusion

If confirmation biases, challenges in data sparsity, and the type of information which to collect from students are not addressed, it will have detrimental effects on attempts to assess and evaluate the effects of recommender systems within South African higher education. If these data collection strategies are not addressed, biases leading to ethical fallacies within this type of research will persist. This study and the purpose of this paper was to outline a data collection framework specific to recommender systems within this context. In South African higher education, the risk of demography based biases may be systemically created, and a subsequent frame-work needed to be created to address this shortcoming. This justification is further emphasized by reducing collection biases and finding optimal ways to assemble information that can infer ideal impacts in the student journey, while not excluding or marginalizing the users. If confirmation biases, challenges in data sparsity, and the type of information to collect from students are not addressed, it will have detrimental effects to institutions, their respective students and society if recommender systems are implemented within this context without the required scholarly engagement. Edizel et al., (2020) justified the inherent biases that exist within recommend-er systems when implemented within society, fortifying societal and systemic biases in a feedback continuum if not addressed. Edizel et al., (2020) further empirically proved that these biases reinforce stereotypes within ethnicity and potentially other societal labels that may be systemically formed. This implies that these processes are complex in nature and require the appropriate engagement in design, evaluation, and implementation in order to mitigate biases within these strategies. Lastly, transparency in fair machine learning occurs at the prediction step of the recommender problem which implies that transparency in system generated feedback needs to be communicated to the academic advisors on the basis of the algorithm design. Within this work, we reviewed a comprehensive evaluation metric for recommender systems data collection strategies within the education domain, and justified the importance of these strategies in terms of evaluation metrics.

## 6    Implications and Future Research

This research outlined a fundamental step that is required in order to establish the groundwork required for data collection strategies in studying recommender systems within higher education. Future research in this area is required in the form of creat-ing comprehensive and robust data collection instruments as well as establish to what extent the data should be de-identified and stored. Once these instruments are in place, subsequent evaluations of these are required in order to establish which filtering method and systems may be the best fit for the recommender problem in this domain. This also means that the data collection strategies need to include in-formation about the student prior, during, and post their education journey.